%% file: main.tex
\documentclass[12pt]{iopart}

\usepackage{harvard}
\citationmode{abbr}
\usepackage{hyperref}
\usepackage{booktabs}
\usepackage{graphicx}

\expandafter\let\csname equation*\endcsname\relax
\expandafter\let\csname endequation*\endcsname\relax
\usepackage{xcolor}         
\usepackage{amsmath}

\usepackage{fancyhdr} 
\usepackage{lineno}

\begin{document}

\title{Combining GEDI and Sentinel-2 for wall-to-wall mapping of tall and short crops}

\author{Stefania Di Tommaso\textsuperscript{1}, Sherrie Wang\textsuperscript{1,2,3}, 
David B. Lobell\textsuperscript{1}{*}}

\address{\textsuperscript{1} Department of Earth System Science and Center on Food Security and the Environment, Stanford University}
\address{\textsuperscript{2} Institute for Computational and Mathematical Engineering, Stanford University}
\address{\textsuperscript{3} Goldman School of Public Policy, University of California, Berkeley}
\ead{dlobell@stanford.edu}

\input{sections/abstract}
\input{sections/intro}

\input{sections/data}

\input{sections/methods}

\input{sections/results}

\input{sections/discussion}

\input{sections/conclusions}
\input{sections/acknowledgements}
\input{sections/appendix}
\input{sections/references}

\end{document}

%% file: sections/abstract.tex
\begin{abstract}
\label{sec:abstract}
High resolution crop type maps are an important tool for improving food security, and remote sensing is increasingly used to create such maps in regions that possess ground truth labels for model training. However, these labels are absent in many regions, and models trained in other regions on typical satellite features, such as those from optical sensors, often exhibit low performance when transferred. Here we explore the use of NASA’s Global Ecosystem Dynamics Investigation (GEDI) spaceborne lidar instrument, combined with Sentinel-2 optical data, for crop type mapping. Using data from three major cropped regions (in China, France, and the United States) we first demonstrate that GEDI energy profiles are capable of reliably distinguishing maize, a crop typically above 2m in height, from crops like rice and soybean that are shorter. We further show that these GEDI profiles provide much more invariant features across geographies compared to spectral and phenological features detected by passive optical sensors. GEDI is able to distinguish maize from other crops within each region with accuracies higher than 84\%, and able to transfer across regions with accuracies higher than 82\% compared to 64\% for transfer of optical features. Finally, we show that GEDI profiles can be used to generate training labels for models based on optical imagery from Sentinel-2, thereby enabling the creation of 10m wall-to-wall maps of tall versus short crops in label-scarce regions. As maize is the second most widely grown crop in the world and often the only tall crop grown within a landscape, we conclude that GEDI offers great promise for improving global crop type maps.

\end{abstract}

%% file: sections/intro.tex
\section{Introduction}\label{intro}

Crop type maps are a crucial step toward estimating crop area, mapping yield, studying local nutritional outcomes, and developing hydrological models \cite{boryan2011monitoring,jin2019smallholder}. Recent years have seen significant progress in remote sensing-based crop type mapping, particularly in high-income countries, with maps now produced in the US \cite{cdl}, Canada \cite{aafc}, much of Europe \cite{defourny2019near,belgiu2018sentinel}, and parts of Asia \cite{you202110m}. While often high in accuracy, the models that produce these maps remain local, in the sense that model application is confined to the region where ground labels exist for crop types. Applying these models outside the region of training sees rapid performance declines \cite{wang2019crop,kluger2021two}, because the models largely use optically-sensed time series as features. These time series, which reflect crop phenology, change from region to region as growing season timing, climate, management practices, soil properties, and crop varieties change. As a result, crop type maps remain elusive in places where ground labels are scarce, which includes the vast majority of low- and middle-income countries.

To date, solutions proposed for creating crop type maps in label-scarce regions include matching satellite time series to crop type profiles \cite{foerster2012crop,belgiu2021phenology}, designing machine learning models that need fewer labels to perform well \cite{jean2019tile2vec,tseng2021learning}, substituting crowdsourced labels in lieu of survey-based labels \cite{wang2020mapping}, and investing more resources to collect ground data in low-income regions \cite{lambert2018estimating,jin2019smallholder,rustowicz2019semantic}. Another potential solution is finding remote sensing features that are invariant to geographic shifts --- in other words, finding a remote sensing modality under which a particular crop type looks the same way everywhere on Earth. So far, such a feature has not been found in multi-spectral imagery at the spectral resolution of MODIS, Landsat, or Sentinel-2, or in radar imagery like that acquired by Sentinel-1, but the ever-growing list of sensors offers new possibilities each year.

The Global Ecosystem Dynamics Investigation (GEDI) is a spaceborne light detection and ranging (lidar) sensor that was launched in late 2018 and installed on the International Space Station \cite{dubayah2020global}. As a lidar waveform instrument, GEDI measures the reflection of a laser beam off of vegetation and the ground surface, with a nominal spatial resolution of 25m. The waveforms are then processed to provide information on surface topography, canopy height, canopy cover, and vertical canopy structure \cite{dubayah2020global}. GEDI was designed with the  goal of improving measures of forest canopy structure, and several recent studies have applied GEDI to this end \cite{schneider_towards_2020,healey2020highly,potapov2021mapping}.  Because spaceborne lidar sensors typically provide only a sparse sampling of the Earth's surface --- during its planned mission GEDI will measure 4\% of the land surface --- lidar data are commonly used as a source of training data for models that estimate forest structure from wall-to-wall imaging sensors, such as Landsat \cite{healey2020highly,potapov2021mapping}, Tandem-X \cite{qi2019forest}, or Sentinel-1 \cite{chen_improved_2021,bruggisser2021potential}. 

Although designed for forest systems, the GEDI measures could also prove useful in cropland systems. In particular, crop height may be a more consistent feature of crops across regions than the spectral and phenological features detected by passive optical sensors. For example, Figure \ref{fig:cropheights}
\input{figures_tex/figure1} displays the distribution of reported crop heights for thousands of varieties of different key species stored in the U.S. National Plant Germplasm System (https://npgsweb.ars-grin.gov/), which contains seed samples from around the world. Among the four staple crops grown most widely across the world, maize is clearly taller than the others, with even the 25th percentile of maize samples exceeding the 95th percentile of the other three crops (rice, wheat, and soybean). On average, maize is roughly 1m taller than the other crops, with 1m equal to the reported vertical resolution of GEDI {\cite{dubayah2020global}}. Thus, it is plausible that GEDI could distinguish maize from other common staples, although it is unlikely that it could distinguish wheat from rice or soybean.  

In this paper, we explore the potential of GEDI to distinguish between taller and shorter crops, and thus to provide more generalize-able features that can be used to transfer crop type models from one region to another. Although several tall crops are commonly cultivated --- Figure \ref{fig:cropheights} illustrates that crops such as sorghum and sunflower exhibit a similar height distribution as maize --- we focus on maize for two main reasons. First, it is by far the most widely grown tall crop in the world, with many regions relying on maize either directly or indirectly (via animal feed) for a substantial portion of their calories and protein. In sub-Saharan Africa, for instance, distinguishing maize from all other crops is often a key step towards estimating national grain supply \cite{jin2019smallholder,nakalembe2021review}. Second, maize is the predominant tall crop in the regions for which we have extensive field-scale crop maps to test our crop type estimates. 

Since we are interested in finding geographically-invariant features, we test GEDI in three maize-producing regions around the world: the state of Iowa in the US, the province of Jilin in China, and the region of Grand Est in France. The three study areas were chosen for their geographic diversity and availability of accurate, up-to-date field-scale crop type maps \cite{cdl,france,you202110m}. By training maize classifiers within each region and applying them across regions, we show (1) GEDI data can distinguish maize from non-maize crops based on height, (2) GEDI features transfer much better than optical features 
across regions spanning multiple continents, and (3) GEDI data can generate training labels that then enable wall-to-wall crop type mapping with optical imagery in the absence of other ground labels.

%% file: figures_tex/figure1.tex
\begin{figure}
	\centering
	\includegraphics[width=1.0\linewidth]{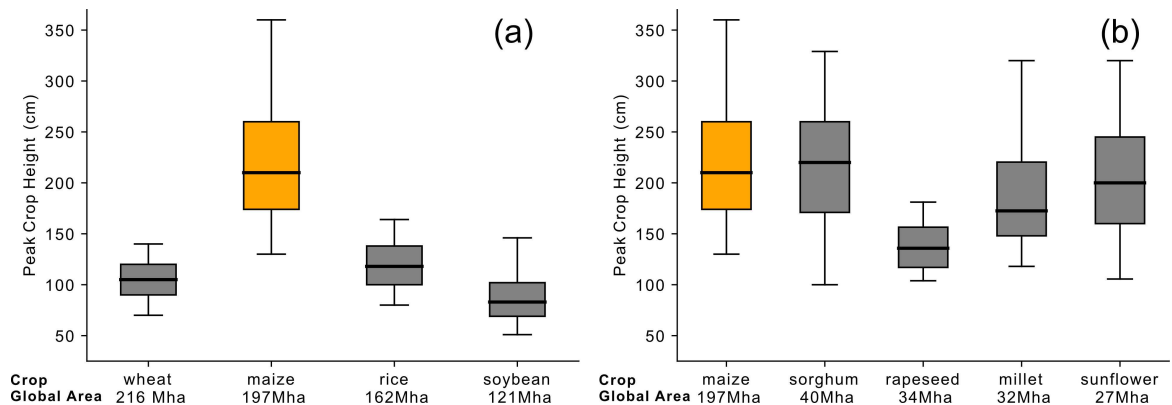}
	\caption{ The distribution of crop heights for the four most widely grown crops in the world (a), as reported for germplasm in the U.S. National Plan Germplasm System. Central lines indicate the median height, boxes indicate the 25th-75th percentile, and whiskers indicate the 5th-95th percentile. Panel (b) compares maize with other tall crops commonly grown in the world. Data source: \url{https://npgsweb.ars-grin.gov} for crop heights, \url{http://www.fao.org/faostat} for global crop areas in 2019.}
	\label{fig:cropheights}
\end{figure}

%% file: sections/data.tex
\section{Datasets}
\label{sec:data}

\subsection{Study areas}
\label{sec:studyarea}

To evaluate the potential of GEDI to distinguish between tall and short crops, we considered three regions of the world: Jilin in  China,  Grand  Est in France, and Iowa in the United States (Figure \ref{fig:studyregion}). 
\input{figures_tex/figure2}
These regions are representative of major agricultural production areas on three separate continents, contain a mix of tall and short crops, and have accurate, up-to-date field-scale crop type maps that are publicly available. 

Jilin Province is located in Northeast China and is one of the most important maize-producing provinces in China. It spans the mid-latitudes from 40.8°N--46.3°N and 121.7°E--131.3°E and has a humid continental climate. Other major crops grown in the area are soybeans and rice. Because early frost usually appears in September and early October, fast maturing maize varieties are cultivated. Maize is typically planted in April and harvested in September, with an average maize cycle duration of 150 days.

The Grand Est administrative region in northeastern France cultivates a wide variety of crops including wheat, barley, maize, alfalfa, sugar beets, legumes, and oilseeds. 
Twenty-two percent of French sugar production, 21\% of rapeseed production, 13\% of wheat production, and 13\% of maize production come from this region \cite{agreste}. Maize is generally planted between April and May, and harvested between September and November. The region spans 47.4°N--50.2°N and 3.4°E--8.2°E and has a climate that varies from oceanic in the west to humid continental in the east. While the administrative region Nouvelle-Aquitaine is the largest producer of maize in France (31\%), Nouvelle-Aquitaine also produces significant quantities of sunflower, which is also a tall plant. To focus on evaluating GEDI's ability to distinguish maize, we conducted experiments in Grand Est, which is France's second-largest producer of maize, instead. We elaborate on the application of GEDI in regions with more than one tall crop in the Discussion.

Iowa, a state located in the Midwestern region of the United States, is in the heart of the U.S. Corn Belt and is the country's largest producer of maize. Maize and soybean are the two primary crops cultivated, comprising well over 95\% of total cropped area (Fig. \ref{fig:crophist}). Maize in Iowa is planted from late April to May and harvested from late September to early November. Located between 40.4°N--43.5°N and 90.1°W--96.6°W, Iowa also experiences a humid continental climate with cold winters and hot summers.

\subsection{GEDI data and feature extraction}
\label{sec:gedi}

The GEDI instrument is the first spaceborne lidar instrument specifically optimized to measure vegetation structure. By firing a laser at 25-meter spots (termed ``footprints'') on the Earth's surface and observing the return of the laser pulse, the instrument is able to measure the vertical distribution of vegetation at each spot. GEDI collects data globally (between 51.6°N and 51.6°S latitudes) at the highest resolution and densest sampling of any lidar instrument in orbit to date \cite{dubayah2020global}. The raw GEDI waveforms collected undergo several processing steps to retrieve a variety of metrics, and the derived products are saved as both footprint and gridded datasets. 

For our analysis, we used the Level 2A Elevation and Height Metrics Data (L2A), which includes footprint-level elevation and relative height (RH) metrics. RH metrics represent the height (in meters) at which a percentile of the laser’s energy is returned relative to the ground. For example, $\text{RH50} = 20 \text{m}$ means that 50\% of the laser's energy was returned by objects up to 20 meters above the ground. The ground position is determined based on the center of the lowest mode of the returned waveform \cite{hofton2000decomposition}. 
RH metrics are saved at 1\% intervals, so each shot contains 101 values representing RH at 0-100\%.
These footprint data are geolocated with a mean positional error of 10.3 m.

We downloaded the GEDI L2A version 2 (\texttt{GEDI02\_A v002}) data from July to September 2019 that intersects our study regions through NASA’s Earthdata Search website.
GEDI L2A data come with a series of flags and properties to help the user filter for data of quality appropriate for the specific application. For this study, we omitted shots with a quality flag value of zero, which indicates poor quality, and a non-zero degrade flag, which indicates poor geolocation. We note that, unlike RH values observed for forests, RH values used here were commonly below zero. This happens because waveforms from agricultural areas often have only one mode, and the GEDI algorithms define RH relative to the center of the lowest mode.
We also filtered out shots with RH100 greater than 10m, as the field crops in the study areas do not grow that tall. We also dropped a full orbit for September 25 in Jilin, China because of abnormally high RH100 values. The outliers removal only dropped a small percentage of points (1.5\% in Iowa, 3.6\% in Jilin, and 3.5\% in Grand Est).
A map of the shots left in each region after cleaning the dataset and filtering for cropland are shown in Fig. \ref{fig:studyregion}, and the counts of shot numbers for each crop type are summarized in Fig. \ref{fig:crophist}.

Consecutive RH metrics are highly correlated with each other. We therefore sampled a metric every 10\% to reduce the number of features used from 101 to 11. The difference in accuracy of random forest classifiers trained on all 101 RH metrics and a subset of 11 is only slightly lower for all three study regions.

This suggests that the information lost from reducing feature dimensionality had little impact on the ability to distinguish crop types.

\subsection{Sentinel-2 imagery and feature extraction}
\label{sec:sentinel2}

Current state-of-the-art crop type maps use imagery from passive optical remote sensing as input features for classification \cite{cdl,defourny2019near}. To compare GEDI features to optical features for crop type classification, we extracted S2 time series at each GEDI shot location. We also extracted S2 time series for the entirety of the study areas in order to demonstrate how GEDI can be used to create labels for wall-to-wall crop type mapping.
All optical imagery was processed using the Google Earth Engine (GEE) platform.

The Sentinel-2A/B (S2) satellites acquire images with a spatial resolution of 10-meters (Blue, Green, Red, and NIR bands) and 20-meters (Red Edge 1, Red Edge 2, Red Edge 3, Red Edge 4, SWIR1, and SWIR2 bands), and together they provide images at a 5-day interval. The spatial resolution of 10-m to 20-m is sufficient to resolve individual fields in the three study areas.

We used S2 surface reflectance data (Level-2A) present in GEE and filtered out clouds using the S2 Cloud Probability dataset provided by SentinelHub in GEE.
To capture crop phenology, we used Sentinel-2 imagery from January 1 to December 31, 2019, using the same time window across the three regions. In our study areas, this time window encompasses a single growing season for the majority of crop types.

Features were extracted from S2 time series by fitting harmonic regressions to all cloud-free observations in 2019. For each spectral band or vegetation index $f(t)$, the harmonic regression takes the form
\begin{equation}
f(t) = c + \sum_{k=1}^{n} \left[ a_{k} \cos (2 \pi \omega k t) + b_{k} \sin(2 \pi \omega k t) \right]
\end{equation}
where $a_{k}$ are cosine coefficients, $b_{k}$ are sine coefficients, and $c$ is the intercept term. The independent variable $t$ represents the time an image is taken within a year expressed as a fraction between 0 (January 1) and 1 (December 31). The number of harmonic terms $n$ and the periodicity of the harmonic basis controlled by $\omega$ are hyperparameters of the regression. We used a second order harmonic ($n=2$) with $\omega=1.5$, shown in previous work \cite{wang2019crop} to result in good features for crop type classification. This yields a total of 5 features per band or vegetation index, resulting in 20 harmonic coefficients total.

We computed harmonic coefficients for three bands and one vegetation index: NIR, SWIR1, SWIR2, and GCVI.
GCVI is the green chlorophyll vegetation index \cite{gitelson2005remote} computed as
\begin{equation*}
\text{GCVI} = \text{NIR} / \text{Green} - 1
\end{equation*}
Unlike the commonly-used NDVI, GCVI does not saturate at high values of leaf area and has previously been shown to aid in distinguishing crop types \cite{wang2019crop}.

\subsection{Crop type labels}
\label{sec:methods_crop_type_labels}

\input{figures_tex/figure3}

We used high-accuracy crop type maps in Jilin, China, Grand Est, France and Iowa, USA to filter out non-crop areas, train maize classifiers, and evaluate each classifier's performance. In each region, the corresponding map's value at each GEDI shot footprint centroid location or S2 pixel location was used as the ground truth for crop type. 
Note that GEDI footprints have a 12.5 m radius, so it is possible for a GEDI shot to span multiple crop type map pixels and have mixed crop type labels (Fig. \ref{fig:adaircounty}). 
The data products available in each region for the year 2019 are described below.

\subsubsection{You \emph{et al.} (2021) Northeast China Crop Type Map}
\label{sec:data_youetal}

You \emph{et al.} (2021) produced annual 10 m crop type maps in Northeast China from 2017 to 2019 for the three major crops in the area (maize, soybean, and rice) using S2 time series data and ground samples from field surveys. 
The overall accuracy for the 2019 crop map for the whole Northeast region is 87\%, with F1-scores of 94\%, 85\%, and 87\% and rice, maize, and soybeans, respectively. Maize and soybean have higher recall (producer's accuracy) (86\% and 90\%) than precision (user's accuracy) (both 84\%), indicating that the commission errors of maize and soybean are higher than the omission errors. According to the authors, this mainly resulted from the incorrect identification of other crops as maize and soybean.

We imported the 2019 crop type map for the province of Jilin in GEE and used it to sample crop type labels at GEDI shot locations for the three major crops mapped.

\subsubsection{Registre Parcellaire Graphique (RPG)}

The Registre Parcellaire Graphique (RPG) is an geographical database of agricultural fields in France maintained by the Service and Payment Agency (ASP). The ASP is the institution that pays aid to French farmers under the Common Agricultural Policy (CAP) of the European Union. As part of their request for CAP aid, farmers send the ASP plot boundaries and certain plot characteristics. Unlike the CDL in the US and the You \emph{et al.} (2021) map in Northeast China, the RPG in France is a georeferenced vector product derived via survey, rather than a raster product generated by a machine learning algorithm. Each plot is drawn to centimeter resolution and associated with a crop type also submitted by the farmer \cite{asp2019rpg}. 

An anonymized version of the dataset is released publicly by the ASP each year, and we accessed this dataset at \url{https://www.data.gouv.fr/}. The entire 2019 database contains 9.6 million plots; filtering for those that fall within the Grand Est region results in a dataset of 851,090 plots. Although the RPG does not include farmland not receiving CAP aid, in reality 98\% of agricultural land in Grand Est is recorded in the RPG.

We imported the RPG dataset in GEE, filtered out non-crop parcels, and rasterized it to sample the crop type labels at GEDI shot locations.

\subsubsection{USDA Cropland Data Layer (CDL)}

Each year, the US Department of Agriculture (USDA) produces the Cropland Data Layer (CDL) for the lower 48 states of the US. A raster product with pixels at 30m resolution, CDL covers 132 classes spanning field crops, tree crops, developed areas, forest, and water. 
It is the output of a decision tree algorithm trained on ground labels obtained through surveys and a combination of Landsat, Disaster Monitoring Constellation, ResourceSat-2, and S2 imagery \cite{cdl_metadata}. The accuracy of CDL labels varies by class and geographic region but is generally high.

We accessed CDL via GEE and used it to filter out GEDI shots in non-crop areas of Iowa and assign crop type labels to GEDI shots in cropped areas. Of the cropped area, 57\% of GEDI shots are maize and 41\% are soybean (Fig. \ref{fig:crophist}). In the 2019 Iowa CDL, maize is classified with a precision of 97\% and recall of 95\%, and soybean is classified with a precision of 96\% and recall of 95\% \cite{cdl_metadata}, indicating that CDL is accurate enough to be used as ground truth to evaluate GEDI features for maize classification.

%% file: figures_tex/figure2.tex
\begin{figure}
	\centering
	\includegraphics[width=1.0\linewidth]{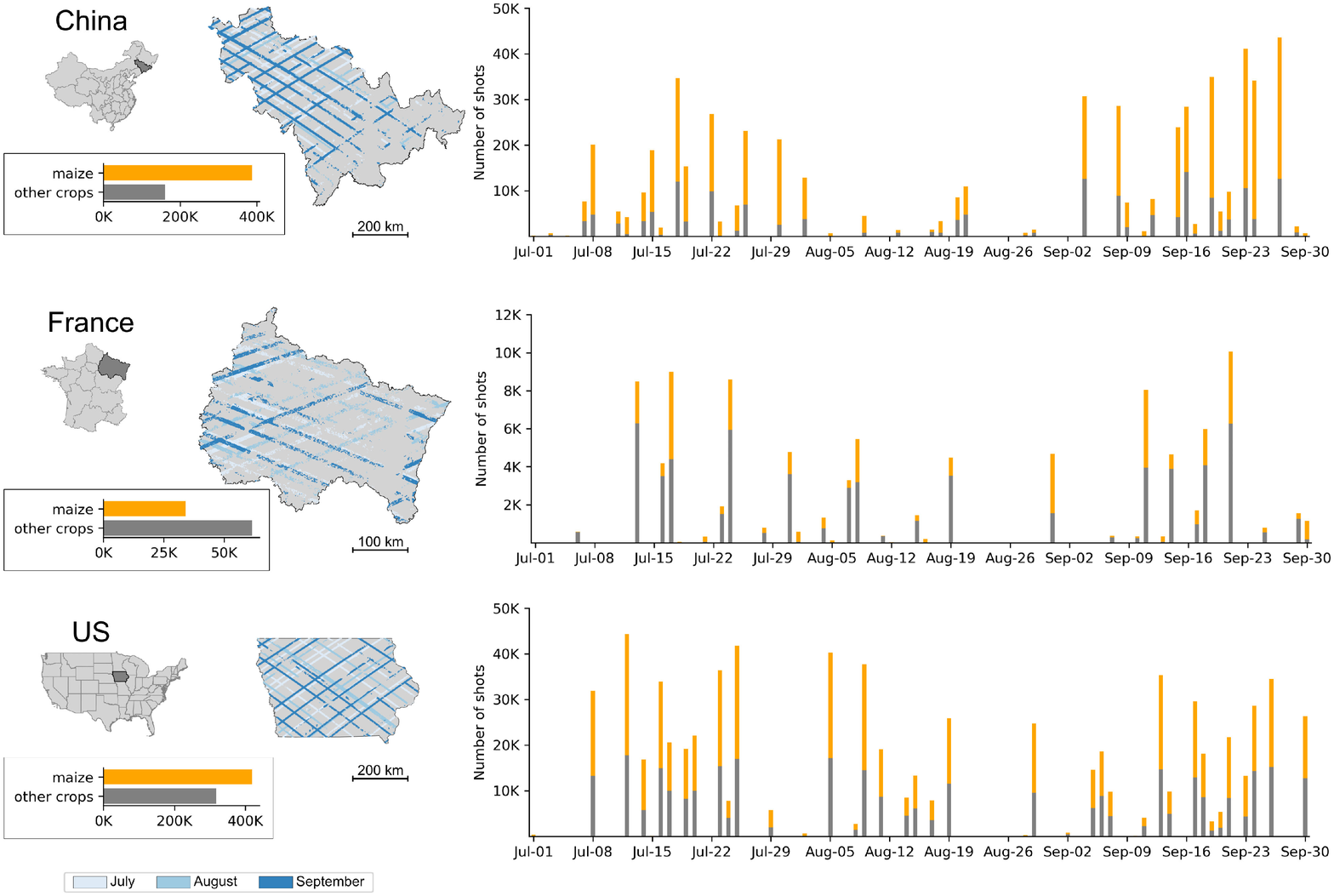}
	\caption{Overview of GEDI observations in the three study regions: Jilin in  China,  Grand  Est in France, and Iowa in the United States. Maps show the shot locations for July-September for cropped area, and barplots indicate the number of shots from maize and non-maize fields for each observation date. }
	\label{fig:studyregion}
\end{figure}

%% file: figures_tex/figure3.tex
\begin{figure}
	\centering
	\includegraphics[width=1.0\linewidth]{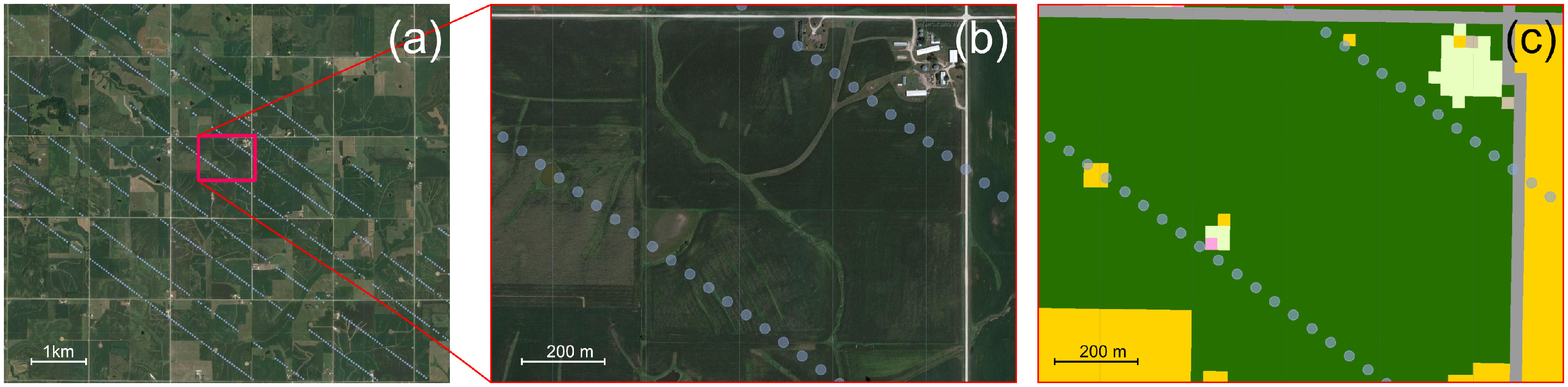}
	\caption{ Example of GEDI shot footprints in cropped area in Adair County, Iowa, U.S. In (a) the eight GEDI laser ground tracks and footprints are separated by 60m along-track and 600m across-track. A close-up of the shots with either a high resolution satellite image as background (b) or the crop type map as background (c) illustrate the typical distribution of shots within fields, including some shots that fall on field boundaries.}
	\label{fig:adaircounty}
\end{figure}

%% file: sections/methods.tex
\section{Methods}
\label{sec:methods}

\subsection{Assembling training and test sets}

For each region, we split the GEDI shot locations into a training (80\%) and test set (20\%) for training and evaluating the models. We discretized each study region into $0.5 \times 0.5$ degree grid cells; all GEDI shots in each grid cell were placed entirely in either the training set or test set. Splitting GEDI shots along grid cells maximizes the chance that samples from the same field are kept together in the train or test set, thereby preventing classification metrics from being inflated due to leakage.

We ran each classification task 11 times using different training and test splits set each time and reported the mean and standard deviation of accuracy over all runs.

\subsection{Random forest classifier}

We used a random forest classifier to classify crop types in all experiments. Random forests \cite{ho1995random,breiman2001random} are an ensemble machine learning method comprised of many decision trees in aggregate. Each decision tree is trained on a bootstrapped version of the training set and a random subset of features to reduce the correlation of predictions across decision trees and improve performance when those predictions are averaged. Random forests are commonly used in crop type classification \cite{defourny2019near,jin2019smallholder} and other Earth observation tasks due to their high accuracy and computational efficiency.

We used the \texttt{RandomForestClassfier} implemented in Python's \texttt{scikit-learn} package. We kept the default parameters, with the exception of raising \texttt{n\_estimators} from 10 to 100 to reduce prediction variance.

\subsection{Crop type classification}
In Table \ref{table:overview} we give an overview of the experiments tested in this paper and detailed below.
\input{tables/table1.tex}

\subsubsection{Testing GEDI features for crop type classification}

Our first experiments test how well GEDI waveforms can distinguish maize from non-maize crops due to maize being a significantly taller crop. As GEDI was designed to monitor forests, it is unknown whether the instrument would be able to resolve height differences between crop types at all.

We trained two random forest classifiers for each study region, one using GEDI RH metrics (GEDI Local) as features, and one using the S2 harmonic coefficients (S2 Local). Both models were tasked with distinguishing maize samples from non-maize samples, where the ground truth for crop type was provided by the datasets described in Section \ref{sec:methods_crop_type_labels}. 
The S2 Local model is representative of current state-of-the-art crop type classifiers that use optical imagery to predict crop types. It provides a reference for the GEDI Local model as well as models that are transferred across regions.

The number of samples and locations used for training and testing the two models were identical. Although S2 provides wall-to-wall imagery while GEDI only observes a small subset of Earth's surface, we limited S2 samples to GEDI shot locations. By controlling for sample size and location, we directly compare the two sets of features.

The timing of GEDI observations is important, as maize will be most distinguishable from other crops when their height difference is the greatest. To test the sensitivity of classification performance to growing season timing, we compared the performance of GEDI Local models trained on July only shots, August only shots, September only shots, and shots from all three months.

\subsubsection{Testing GEDI feature transfer across regions}

After classifying maize versus non-maize crops within each region, we tested model transfer across regions.
For each GEDI Local and S2 Local classifier trained in the U.S., China, or France, we applied the classifier to the test sets of the other two regions to separate maize from non-maize. We refer to these models as GEDI Transfer and S2 Transfer. The models were not shown any additional data from the new regions. High classification accuracy in a new region would indicate that the model's features generalize across space and few if any labels are needed from the new region to classify maize; conversely, low classification accuracy would mean that the learned relationship between features and crop types holds true only locally, and labeled data is needed in the new region to learn new classification boundaries.

We compared the GEDI Transfer models to their Local counterparts to see how well GEDI RH metrics generalize across geography. To understand whether growing season timing affects model transfer, we repeated this analysis for each GEDI model trained on July only shots, August only shots, September only shots, and shots from all three months.

\subsubsection{Wall-to-wall crop type mapping using GEDI as training labels}

Even if GEDI features transfer perfectly across geography---i.e. maize is always identifiable in GEDI waveforms no matter where on Earth one looks---GEDI only samples 4\% of the land surface and cannot alone generate a wall-to-wall crop type map. Achieving a continuous map in space requires GEDI-based approaches to be combined with wall-to-wall imagery like that provided by S2.

To create wall-to-wall maize maps using GEDI and S2 imagery, we used methods similar to those employed previously to calibrate local maps of forest height with GEDI and Landsat \cite{healey2020highly}.
In a new region, we trained a model (GEDI-S2 Transfer) using predicted crop types from GEDI Transfer as labels and local S2 harmonics as features. By applying the GEDI-S2 Transfer model to all cropland pixels in a new region, we produced a wall-to-wall 10m spatial resolution maize map without the need for local labels. Figure \ref{fig:workflow} presents a graphical explanation of the GEDI-S2 Transfer approach.
\input{figures_tex/figure4}

We compared GEDI-S2 Transfer to two other models. The first is the S2 Local model, which provides an upper bound for how well S2 harmonic features can classify maize when trained on in-region ground truth. The gap between GEDI-S2 Transfer and S2 Local reflects the accuracy of GEDI Transfer's predictions relative to ground truth. The second benchmark is the S2 Transfer model, which shows how well a model trained on S2 harmonics in one region fares when applied to other regions. Differences between S2 Transfer and GEDI-S2 Transfer reveal how robust GEDI features are across space compared to optical features.

%% file: tables/table1.tex
\begin{table}[t]
\footnotesize
\centering
\begin{tabular}{p{0.15\linewidth} p{0.2\linewidth} p{0.13\linewidth} p{0.15\linewidth} p{0.1\linewidth} p{0.07\linewidth}}
\toprule
\textbf{Method} & \textbf{Description} & \textbf{Features} & \textbf{Training} & \textbf{Spatial} & \textbf{Local} \\ 
\textbf{Name} &  &  & \textbf{Labels} & \textbf{Coverage} & \textbf{Labels} \\
\midrule

S2 Local 
&    
Training and tested in different locations of the same region
&
Sentinel-2 \newline (20 harmonic coefficients)
&
Training region crop type map
& 
wall-to-wall 
& 
yes \\ \hline

GEDI Local 
&
Training and tested in different locations of the same region
&
GEDI \newline (11 RH \newline metrics)
&
Training region crop type map
& 
point 
& 
yes \\ \hline

S2 Transfer 
&    
Training and tested in different regions
&
Sentinel-2 \newline (20 harmonic coefficients)
&
Training region crop type map
& 
wall-to-wall 
& 
no \\ \hline

GEDI Transfer 
&
Training and tested in different regions
&
GEDI \newline (11 RH \newline metrics)
&
Training region crop type map
& 
point 
& 
no \\ \hline

GEDI-S2 Transfer 
&    
Training and tested in different locations of the same region
&
Sentinel-2  \newline (20 harmonic coefficients)
&
GEDI Transfer predictions
& 
wall-to-wall 
& 
no \\


\bottomrule
\end{tabular}
\vspace{10pt}
\caption{Summary of the experiments.}
\label{table:overview}
\end{table}

%% file: figures_tex/figure4.tex
\begin{figure}
	\centering
	\includegraphics[width=1.0\linewidth]{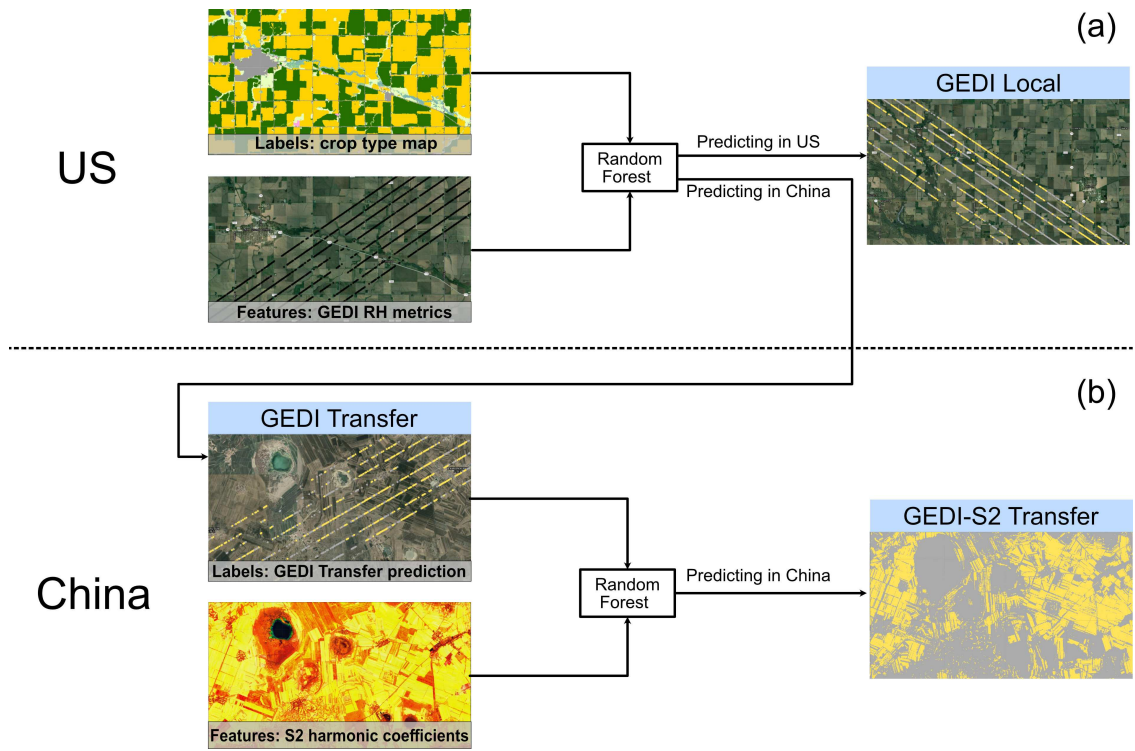}
	\caption{ Graphical overview of the GEDI-S2 Transfer approach: an example of transferring a model trained on US crop type labels and applying it to China to create wall-to-wall maize maps. 
	In (a) a random forest model is trained using GEDI RH metrics as features and CDL as labels in Iowa, U.S. The model is tested in a different region within Iowa (GEDI Local) to evaluate model performance.
	In (b) the Iowa-trained model is used to create predictions in China at GEDI shot locations (GEDI Transfer). These predictions then serve as labels for a new model trained on Sentinel-2 harmonic coefficients. Combining Sentinel-2 and GEDI (GEDI-S2 Transfer) enables a wall-to-wall maize map in China without any ground truth labels from China. In this schematic, the GEDI-S2 Transfer predicted map is shown for the same region used for training, but model evaluation is always done using a test set not used in training. }
	\label{fig:workflow}
\end{figure}

%% file: sections/results.tex
\section{Results}\label{results}

\subsection{GEDI and Sentinel-2 feature comparison}
\label{sec:results_features}

The median harmonics of GCVI from S2 and median RH energy curves from GEDI are shown for the top three crops in each region in Fig. \ref{fig:profiles}.
\input{figures_tex/figure5}
In each region, the S2 maize profile reaches peak greenness in August, and is generally distinguishable from the other crops because of different timing and magnitude of the greenness peaks. For the GEDI energy profiles, roughly 50\% of the energy returned comes from negative RH values, which as mentioned previously arises from the GEDI algorithm's definition of ground elevation based on the the center of the lowest mode, with waveforms from cropped areas typically having only one mode. Thus, we emphasize that the values of RH should not be interpreted as physically meaningful; for instance RH100 does not correspond to the physical crop height. Nonetheless, the curves exhibit a clear separation between maize---the tallest crop---and the other crops. This difference is especially apparent at the extremes of RH curves, shown as insets in Fig. \ref{fig:profiles}. 

Whereas Fig. \ref{fig:profiles} compares different crops within a region, Fig. \ref{fig:maizeprofiles} displays the median features for maize from different regions on the same plot. 
\input{figures_tex/figure6}
This comparison is especially relevant for the question of how well a model is likely to transfer across regions. Ideally, features would be similar across regions in order to use a model trained in one region on another. For the S2 harmonics, clear differences emerge between the regions, with the U.S. generally having a steeper increase in GCVI during June and July and a higher peak in August compared to other regions. The harmonic curves for the other crops also differ considerably, both in timing and magnitude of the peak (Fig. \ref{fig:profiles}). In contrast, the GEDI curves are remarkably consistent across regions (Fig. \ref{fig:maizeprofiles}). This difference between S2 and GEDI indicates that the peak height 

of maize is a better preserved characteristic across regions than the timing of the maize growing season and the total crop biomass, both of which influence the S2 harmonics.

\subsection{Local maize classification}
\label{sec:results_local}

\input{figures_tex/figure7}

The feature comparisons above suggest that GEDI features should be reliable for classifying tall crops (in this case, maize). To quantitatively evaluate its performance for local classification, we used these GEDI features to train a classifier for each region and compared it to a classifier trained on the S2 harmonics. 

In all cases, the task is to separate maize from other crops, and we considered various possible timings of the GEDI features. Test accuracies indicate that GEDI metrics can distinguish maize from other crop within region with more than 79\% accuracy in all cases (Fig. \ref{fig:local}).
The optimal timing of GEDI observations differs by region, with September for China, July for France, and August for the U.S. being the best times for classification, resulting in 88\%, 85\%, and 91\% accuracy, respectively. August is generally a good month for GEDI observations in all regions, with performance in all regions above 83\% for this month.

As expected, the locally-trained S2 models do well in each region (93\% accuracy in China, 95\% in France, and 95\% in the U.S.), and indeed perform better than the GEDI features. The gap between S2 local models (trained on the same locations as the July--September GEDI shots) and the best GEDI models is less than 5\% for China and the U.S. and less than 10\% in France.

To understand why GEDI model errors are larger, we show maps of GEDI misclassifications in Appendix Fig. \ref{fig:errormap} and confusion matrices for representative median runs in Appendix Fig. \ref{fig:confusionmatrix}. From the maps, we see that a significant percent of errors occur at field borders, where the GEDI shot footprint contains multiple crop types or a mix of crop and non-crop classes. 
We also observe that the most difficult crop types for GEDI Local models to classify are soybeans in China and silage corn and sunflower in France. In China, this can be partly explained by the 84\% precision for soybeans in the You \emph{et al.} (2021) map used for ground truth (Section \ref{sec:data_youetal}). In other words, up to half of the ``misclassification'' of soybeans as maize in China could be correct. Since the ``ground truth" maps in China and the U.S. are created using optical satellites, the predictions of S2 Local models---both correct and incorrect---are likely to correlate with the ground truth more than those of GEDI Local models.

Local GEDI accuracies in France are overall lower than the other two regions. This appears to arise mainly from the fact that two kinds of maize are grown in France, maize for grain and for silage (see Fig. \ref{fig:crophist} for the crop distribution by region). Whereas grain maize is always grown to maturity, and thus has a more reliable seasonality, silage maize is grown for biomass and thus can be sown and harvested at any point in the season. As a sensitivity test, we recalculate local GEDI accuracy in France after omitting the silage maize from the test set, finding that accuracies improve by 4\% or more in all periods (see Fig. \ref{fig:nosilage} in Appendix). The improvement is largest in September, consistent with the notion that early harvest of silage maize is affecting the performance of GEDI features, since a harvested crop is no longer tall. As both the U.S. and China grow predominantly grain maize, the less predictable timing of silage maize is not an issue in those regions.

\subsection{Transferring classification across regions}
\label{sec:results_transfer}

As noted in Section \ref{sec:results_features}, the consistency of GEDI features across regions suggests that models trained in one region can be reliably applied to new regions. A quantitative test of this proposition is shown in Figure \ref{fig:transfer}, which compares the performance of GEDI models trained using data from the local region (GEDI-Local) to those trained in other regions (GEDI-Transfer). 
\input{figures_tex/figure8}
Although the locally trained models are typically the best performers, the transferred models perform nearly as well and are occasionally indistinguishable from the locally trained models. For example, models trained in the U.S. or China both perform as well in France in August as one trained in France.

Transferred GEDI models are only able to make predictions at GEDI shot locations. In order to extrapolate beyond these point locations, we used transferred GEDI predictions as labels to train a new model that takes local S2 harmonics as input (GEDI-S2 Transfer). GEDI-S2 Transfer test accuracies for the month of August are reported in Fig. \ref{fig:gedis2transfer} \input{figures_tex/figure9} together with S2 Transfer and S2 Local accuracies for comparison. 
S2 Transfer shows how well state-of-the-art optical features perform out-of-region, while S2 Local provides an upper bound on how well optical features can separate maize when paired with ample local ground truth.

The results in Fig. \ref{fig:gedis2transfer} show first that harmonic features, while good at distinguishing crop types locally, transfer poorly across study regions. The average accuracy of an S2 Transfer classifier is 64\%. In the U.S., the S2 Local model achieves an accuracy of 94\%, but S2 Transfer models trained in China and France only manage accuracies of 60\% and 62\%, respectively. This is consistent with previous work that found optical feature transfer-ability to deteriorate across geography \cite{wang2019crop}, as well as with the differences in crop phenology observed in Fig. \ref{fig:maizeprofiles}. As the phenology and prevalence of crop types shift across regions, the optimal decision boundary for classifying maize versus non-maize with harmonic features also changes. S2 Transfer therefore results in many misclassified samples.

GEDI RH features, on the other hand, transfer much better across regions, and consequently the S2 model they supervise (GEDI-S2 Transfer) also performs much better than direct S2 Transfer. 
 
Fig. \ref{fig:gedis2transfer} shows that GEDI-S2 Transfer accuracies exceed 82\% for all cross-region pairs.
For example, GEDI-S2 Transfer 
achieves 86\% accuracy in the U.S. for models trained in China or France. While the S2 Local model has an 8\% higher accuracy, the GEDI-S2 Transfer significantly outperforms S2 Transfer from China and France by 26\% and 24\%, respectively.
GEDI-S2 Transfer and GEDI Transfer accuracies are about the same; in China they are the same, in the U.S. GEDI-S2 Transfer is 1\% lower, and in France GEDI-S2 Transfer is 1\% lower for the model trained in China and 1\% higher for the model trained in the U.S.

Example crop type predictions for the three regions using the best GEDI-S2 Transfer model are shown in Fig. \ref{fig:maps}. The corresponding ground truth crop type maps are also shown for reference.

\input{figures_tex/figure10}

%% file: figures_tex/figure5.tex
\begin{figure}
	\centering
	\includegraphics[width=1.0\linewidth]{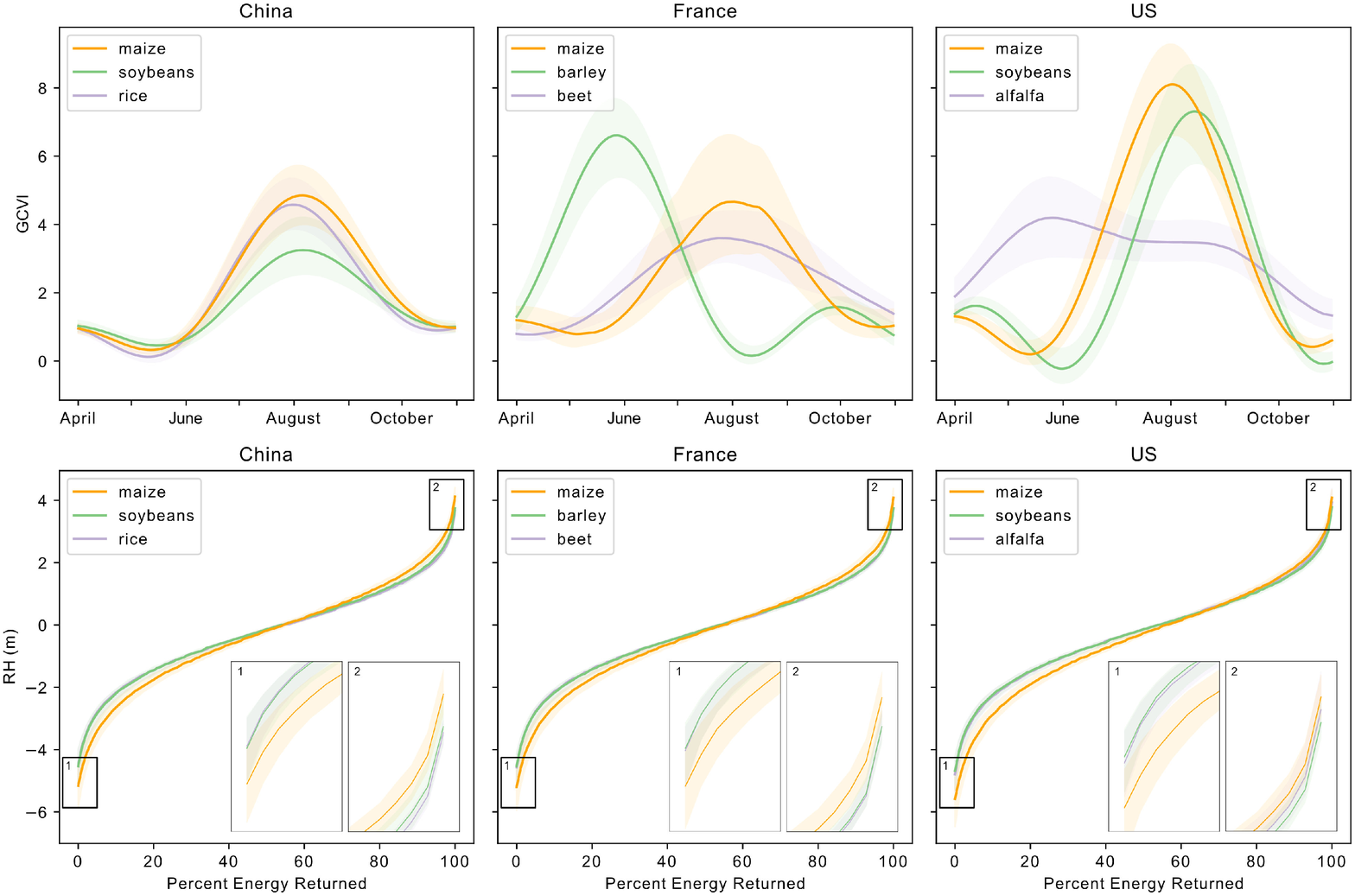}
	\caption{Median harmonics of GCVI from Sentinel-2 (top row) and median RH energy curves from GEDI (bottom row) are shown for the top three crops in each region. GEDI profiles are computed from all shots covering the July-September period. Shading shows the 25th-75th percentile of observations.}
	\label{fig:profiles}
\end{figure}

%% file: figures_tex/figure6.tex
\begin{figure}
	\centering
	\includegraphics[width=0.8\linewidth]{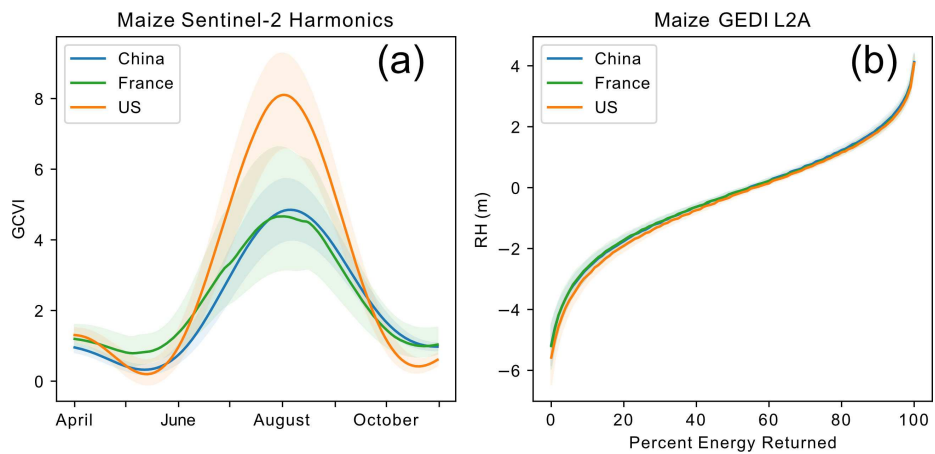}
	\caption{Median maize profiles in the three regions, shown for Sentinel-2 GCVI harmonics in (a) and GEDI curves in (b). Shading shows the 25th-75th percentile of observations.}
	\label{fig:maizeprofiles}
\end{figure}

%% file: figures_tex/figure7.tex
\begin{figure}
	\centering
	\includegraphics[width=1.0\linewidth]{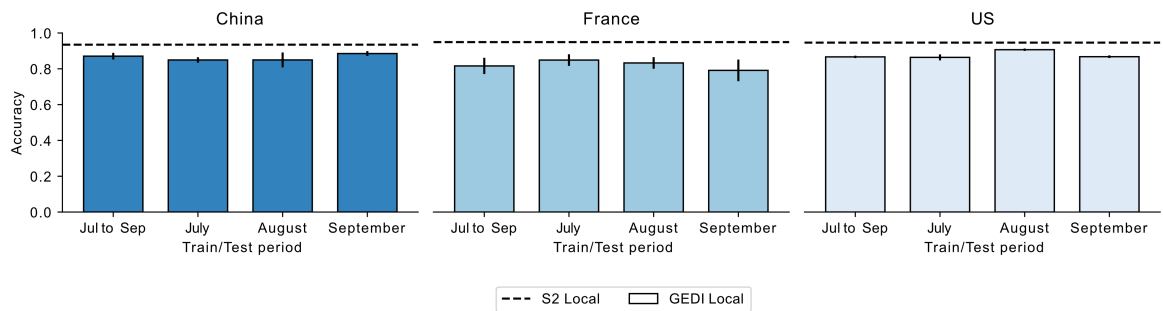}
	\caption{Classification accuracies for locally-trained GEDI and Sentinel-2 models. Bars indicate mean GEDI accuracies for models trained in different months. Error bars show one standard deviation. The dashed line indicates accuracy of the S2 Local model in each region, which is 93\% in China, 95\% in France, and 95\% in the U.S. Note that training sample locations for the Sentinel-2 model were the same as the training shot locations for GEDI for all three months.}
	\label{fig:local}
\end{figure}

%% file: figures_tex/figure8.tex
\begin{figure}
	\centering
	\includegraphics[width=1.0\linewidth]{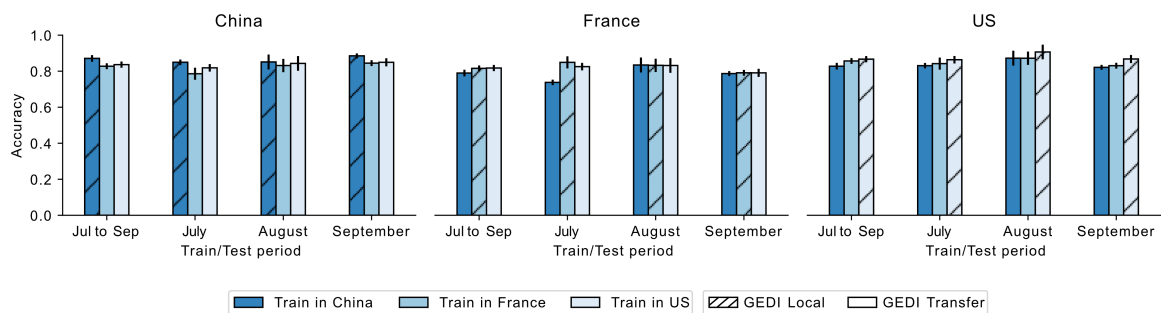}
	\caption{Test accuracies for GEDI models when using different combinations of training and test regions and periods. Colors indicate the region where models were trained, with hatching indicating a model trained in the same region (but on different locations). Models trained in other regions typically perform similarly to those trained locally.}
	\label{fig:transfer}
\end{figure}

%% file: figures_tex/figure9.tex
\begin{figure}
	\centering
	\includegraphics[width=1.0\linewidth]{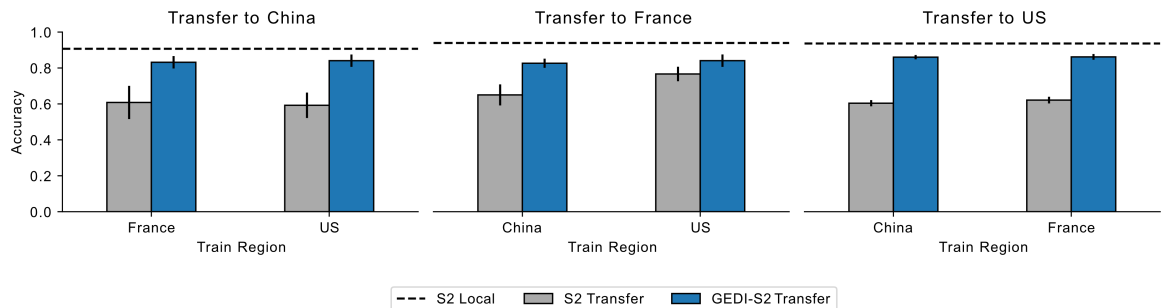}
	\caption{Test accuracies of models using Sentinel-2 harmonic features for wall-to-wall mapping of maize and non-maize for the month of August, trained either with direct transfer of Sentinel-2 features from other regions (gray bars) or using labels from GEDI predictions (blue bars). Error bars show one standard deviation. Dashed lines show performance of a locally-trained Sentinel-2 model (with training samples from the same GEDI shot locations) for the month of August as a comparison.}
	\label{fig:gedis2transfer}
\end{figure}

%% file: figures_tex/figure10.tex
\begin{figure}
	\centering
	\includegraphics[width=1.0\linewidth]{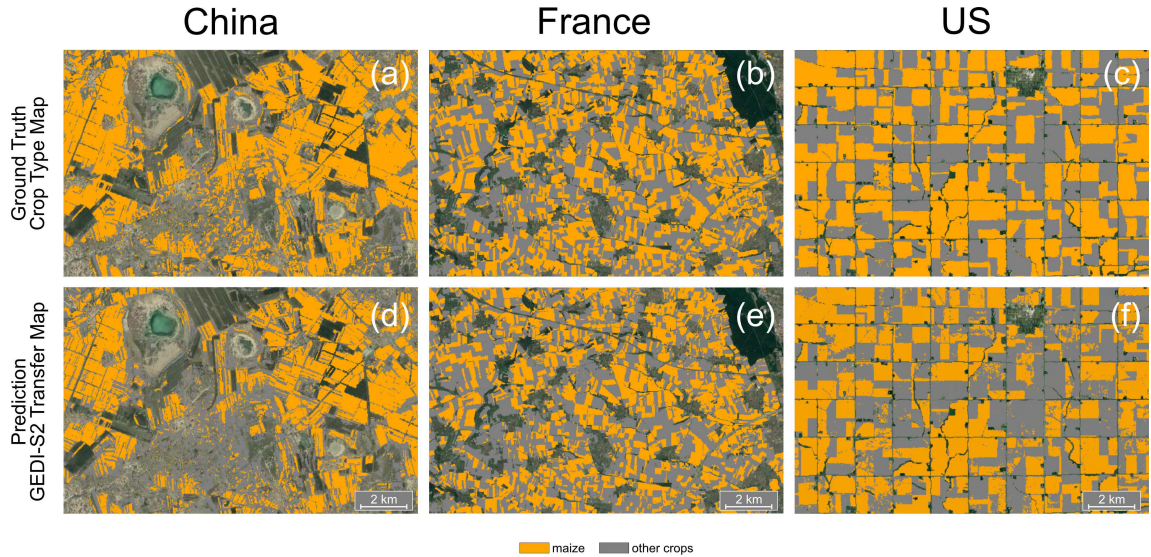}
	\caption{Ground truth crop type maps (a-c) compared with classification predictions of GEDI-S2 Transfer models (d-f). Prediction maps for China (d) and France (e) are created using models trained in the U.S. (84\% accuracy for both). The prediction map for the U.S. was created using the model trained in France (86\% accuracy).}
	\label{fig:maps}
\end{figure}

%% file: sections/discussion.tex
\section{Discussion}\label{discussion}

The results show that GEDI features can distinguish a tall crop like maize from shorter crops, and that these features are highly transferable across geography. 
Should spaceborne lidar sensors someday sample the Earth's surface more densely, they would add a useful set of features for mapping crop types that are complementary to optical and radar features. 
In regions where crop type maps are already available, such as the study areas considered here, lidar could augment field surveys to generate crop type labels, reducing the cost of creating products like CDL. This is especially true in a system like the U.S. Corn Belt, where agriculture is heavily dominated by maize (a tall crop) and soybeans (a short crop). 

Most importantly, lidar has the potential to enable mapping of tall crops like maize in areas of the world where crop type maps are not available due to a lack of ground labels. Our experiments transferring GEDI features and using GEDI to train wall-to-wall crop type maps in China, France, and the U.S. show the robustness of lidar features across continents, despite the GEDI instrument being designed to monitor forests rather than cropland. In fact, S2 models trained on GEDI performed only slightly worse than those trained on local ground truth (Fig. \ref{fig:gedis2transfer}). Although the planned lifetime for GEDI is only two years, the S2 models trained on GEDI could be applied in years without GEDI observations, likely with adjustments to account for potential shifts in features over time \cite{kluger2021two}. 

Despite the promising results, we recognize many potential issues that would emerge when extending this approach to the global scale. First, mapping locations of tall and short crops will not suffice for many applications, which require more detailed crop information. Where maize is the predominant tall crop, as was the case for the three regions studied here, a map of tall crops can be reliably used to identify maize areas. In regions with multiple tall crops, additional features would be needed to separate individual crop types; for instance, optical data has proven useful for distinguishing maize from sorghum \cite{soler2021maize} and radar data for distinguishing maize from sunflower \cite{veloso2017understanding,belgiu2018sentinel}.

Second, many of the errors we observed in GEDI predictions occurred at the edges of fields. These disparities likely reflect some combination of errors in the labels as well as mixed crop types within the GEDI shot footprint. For applications in smallholder regions, these mixed pixels will be increasingly common. It is possible that such errors would have only minimal effect on model training, since they are likely to be random, or that maps of field boundaries could be used to filter out shots near field edges \cite{waldner2020deep}. 

Another issue, particularly in tropical systems, could be frequent cloud cover during the time of year when crops are at peak height, which could limit the availability of clear GEDI shots. Although we observed lower availability of clear shots during August for the current study regions (Fig. \ref{fig:studyregion}), it did not appear to compromise the performance of the model relative to other months with more observations, presumably because many thousands of clear observations were obtained even in the cloudier months. Nonetheless, clouds could emerge as an important constraint in other locations.

%% file: sections/conclusions.tex
\section{Conclusions}
\label{conclusions}

We conclude that GEDI holds great promise for improving agricultural monitoring, because it captures features that are much more generalize-able than those typically used in satellite-based crop type mapping. The demonstrated ability to distinguish crops with height differences of just one meter suggest other potential applications in agriculture should be feasible, such as monitoring the age and growth of tree crops or identifying intercropped fields that contain mixtures of different crops. 

%% file: sections/acknowledgements.tex
\section*{Acknowledgements}

This work was supported by the NASA Harvest Consortium
(NASA Applied Sciences Grant No. 80NSSC17K0652, sub-award 54308-Z6059203 to DBL). Work by SW was partially supported by the Ciriacy-Wantrup Postdoctoral Fellowship at the University of California, Berkeley. We thank the Google Earth Engine team for making large-scale computational resources available to researchers. 

%% file: sections/appendix.tex
\appendix
\section*{Appendix}

\setcounter{figure}{0}

\renewcommand{\thefigure}{A\arabic{figure}}

\input{figures_tex/figureA3}

\input{figures_tex/figureA1}

\input{figures_tex/figureA2}

\input{figures_tex/figureA4}

\clearpage

%% file: figures_tex/figureA3.tex
\begin{figure}
	\centering
	\includegraphics[width=1.0\linewidth]{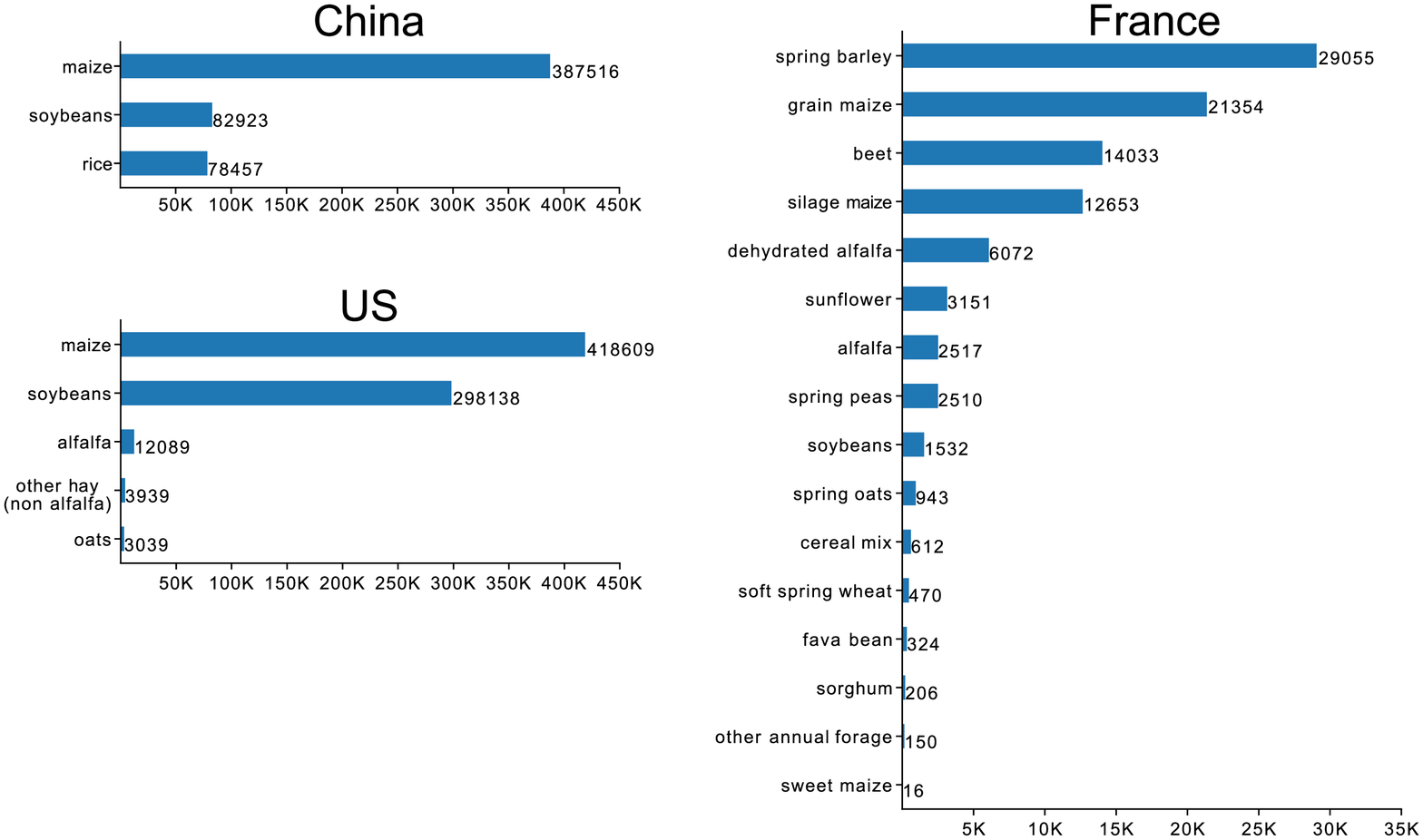}
	\caption{Number of GEDI shots in each region by crop type.}
	\label{fig:crophist}
\end{figure}

%% file: figures_tex/figureA1.tex
\begin{figure}
	\centering
	\includegraphics[width=0.6\linewidth]{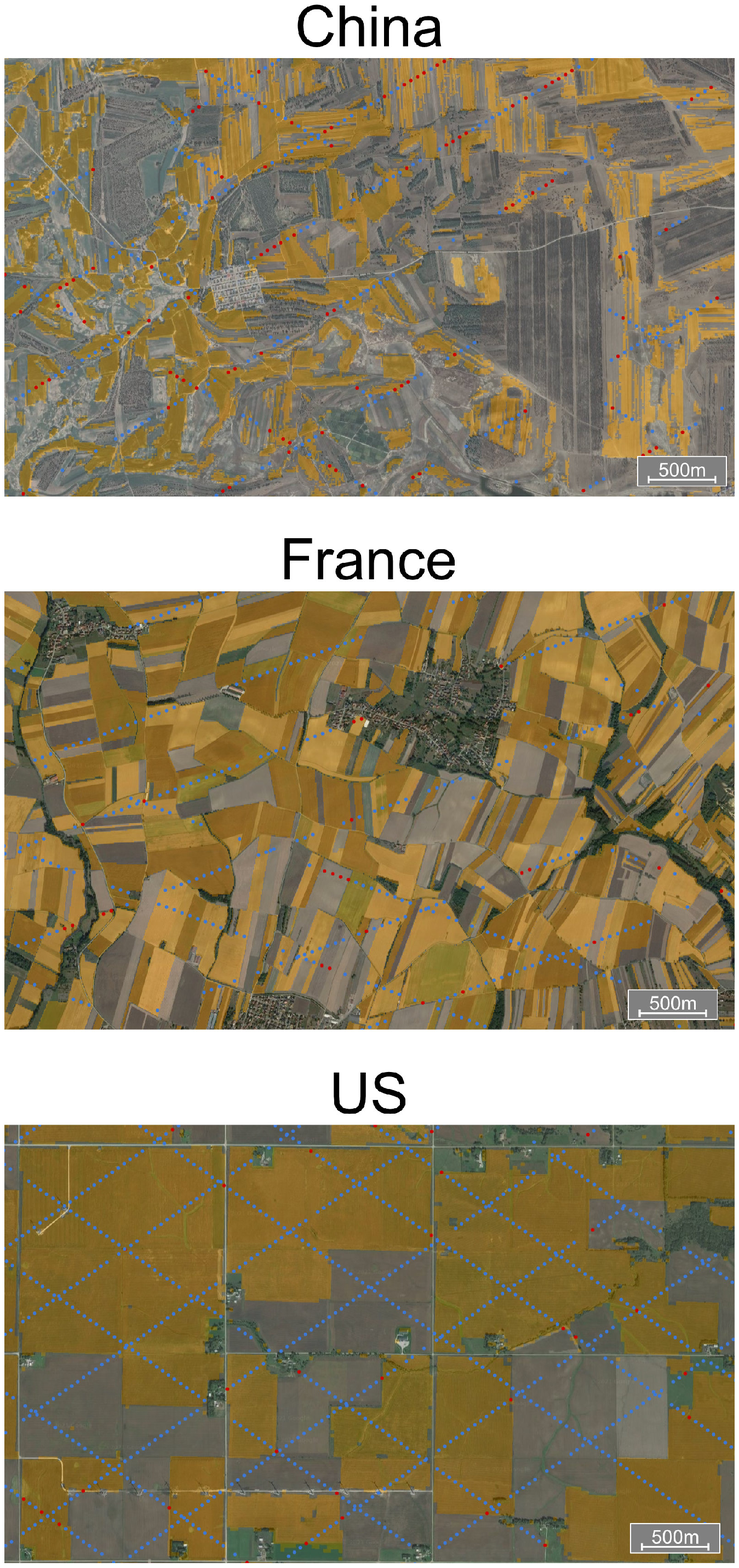}
	\caption{GEDI Local predictions in the three regions. In blue are the correctly classified shots, in red the misclassified ones. Many errors occur at the borders of the fields, likely because of mixed crop types within GEDI shots footprints. Some other errors can be attributed to errors in the crop type maps used as ground truth.}
	\label{fig:errormap}
\end{figure}

%% file: figures_tex/figureA2.tex
\begin{figure}
	\centering
	\includegraphics[width=1.0\linewidth]{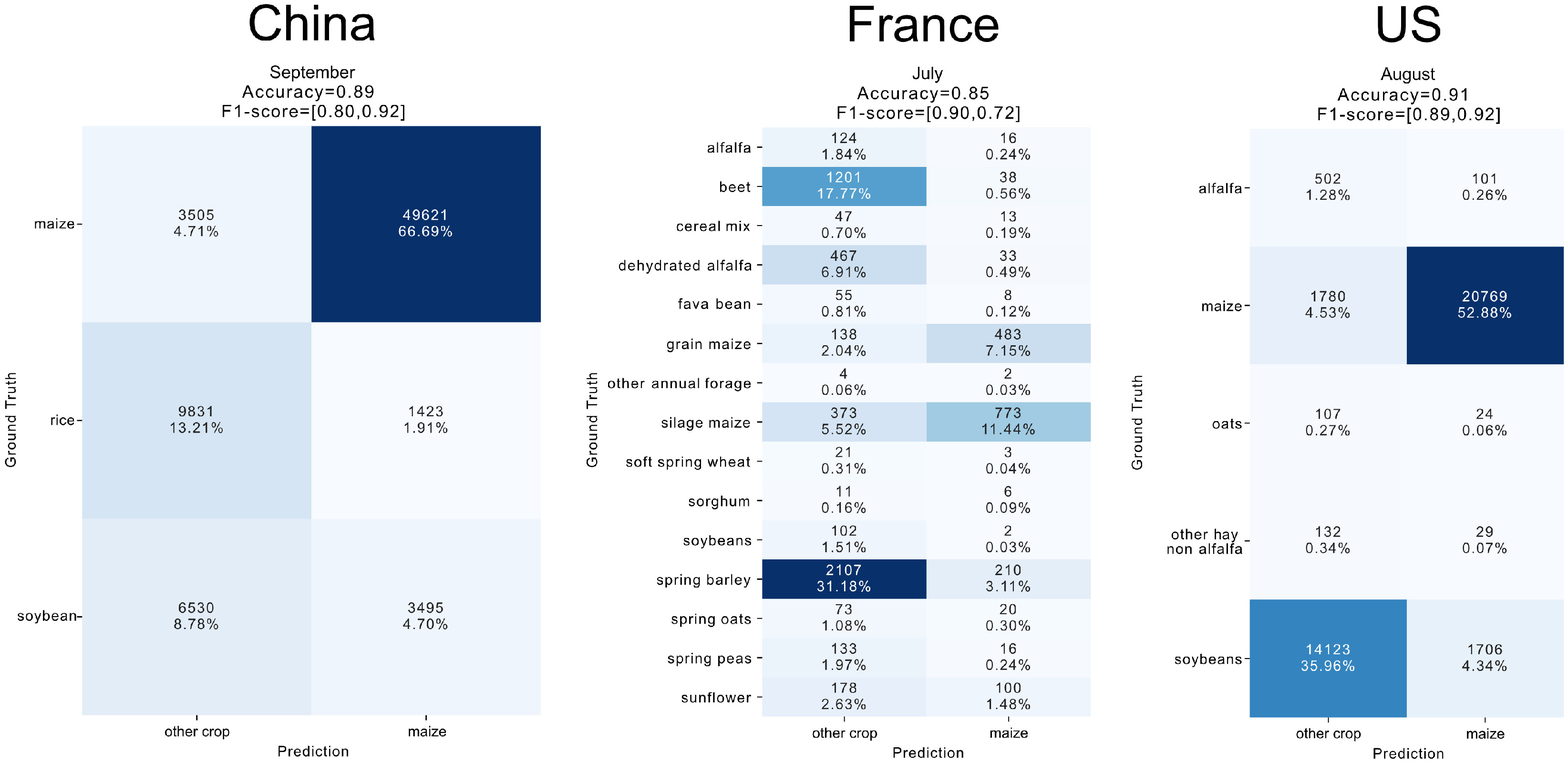}
	\caption{Confusion matrices of GEDI Local classification for the best time in the three regions: September for China, July for France, and August for the U.S., which resulted in a mean accuracies of 88\%, 85\%, and 91\%, respectively. For our analysis, we ran the classification task multiple times, each time with different train and test sets, and computed mean accuracies. Here we are showing confusion matrices only for the median run, i.e. the run with median accuracy. We ran a binary classification (maize vs. other crop). Here we show ground truth labels detailed by crop type to get deeper insights into misclassifications.}
	\label{fig:confusionmatrix}
\end{figure}

%% file: figures_tex/figureA4.tex
\begin{figure}
	\centering
	\includegraphics[width=1.0\linewidth]{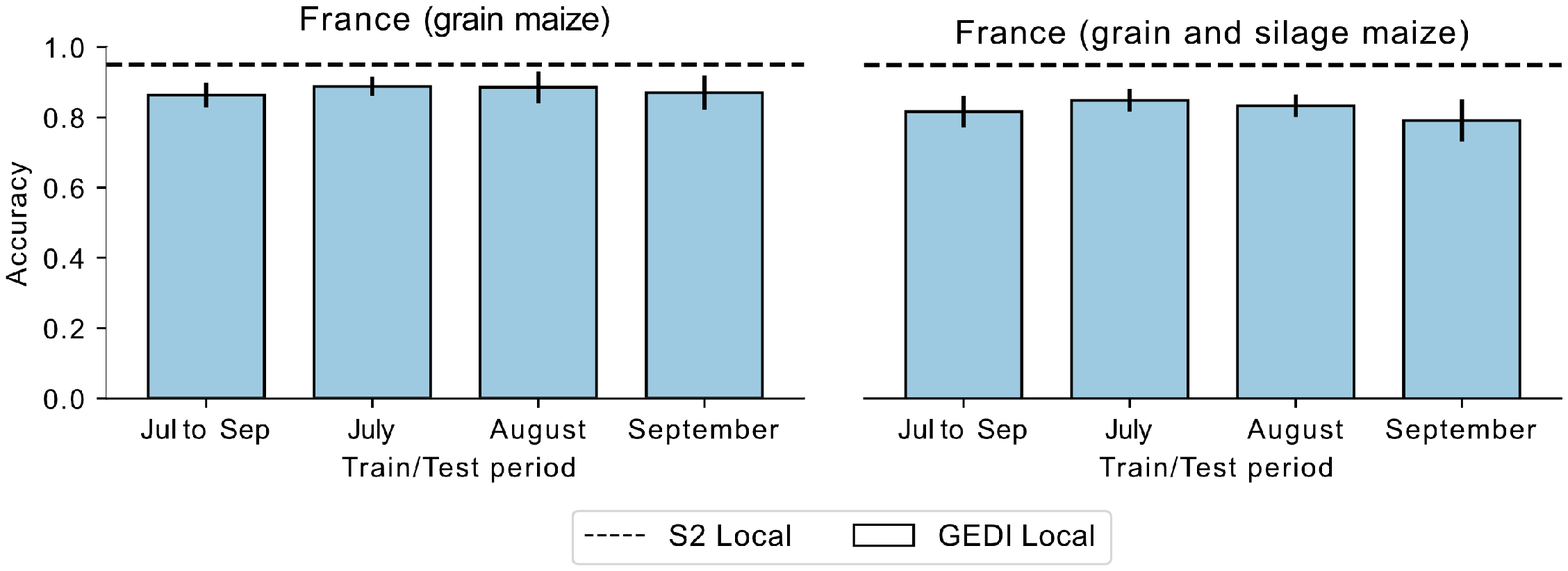}
	\caption{Comparison of GEDI Local accuracies in France when (left) excluding versus (right) including maize for silage. Accuracies improve when considering only grain maize by more than 4\% in all periods, with September improving 10\% most likely due to early harvest of silage maize.}
	\label{fig:nosilage}
\end{figure}

%% file: sections/references.tex
\section*{References}

\bibliographystyle{dcu}
\bibliography{bibl}

%% file: bibl.bib
@article{you202110m,
	title = {The 10-m crop type maps in {Northeast} {China} during 2017–2019},
	volume = {8},
	copyright = {2021 The Author(s)},
	issn = {2052-4463},
	number = {1},
	journal = {Scientific Data},
	author = {You, Nanshan and Dong, Jinwei and Huang, Jianxi and Du, Guoming and Zhang, Geli and He, Yingli and Yang, Tong and Di, Yuanyuan and Xiao, Xiangming},
	month = feb,
	year = {2021},
	note = {Number: 1
Publisher: Nature Publishing Group},
	pages = {41}
}

@article{dubayah2020global,
	title = {The {Global} {Ecosystem} {Dynamics} {Investigation}: {High}-resolution laser ranging of the {Earth}’s forests and topography},
	volume = {1},
	issn = {2666-0172},
	shorttitle = {The {Global} {Ecosystem} {Dynamics} {Investigation}},
	journal = {Science of Remote Sensing},
	author = {Dubayah, Ralph and Blair, James Bryan and Goetz, Scott and Fatoyinbo, Lola and Hansen, Matthew and Healey, Sean and Hofton, Michelle and Hurtt, George and Kellner, James and Luthcke, Scott and Armston, John and Tang, Hao and Duncanson, Laura and Hancock, Steven and Jantz, Patrick and Marselis, Suzanne and Patterson, Paul L. and Qi, Wenlu and Silva, Carlos},
	month = jun,
	year = {2020},
	pages = {100002}
}

@article{schneider_towards_2020,
	title = {Towards mapping the diversity of canopy structure from space with {GEDI}},
	volume = {15},
	issn = {1748-9326},
	number = {11},
	journal = {Environmental Research Letters},
	author = {Schneider, Fabian D. and Ferraz, António and Hancock, Steven and Duncanson, Laura I. and Dubayah, Ralph O. and Pavlick, Ryan P. and Schimel, David S.},
	month = oct,
	year = {2020},
	note = {Publisher: IOP Publishing},
	pages = {115006}
}

@article{chen_improved_2021,
	title = {Improved estimation of forest stand volume by the integration of {GEDI} {LiDAR} data and multi-sensor imagery in the {Changbai} {Mountains} {Mixed} forests {Ecoregion} ({CMMFE}), northeast {China}},
	volume = {100},
	issn = {0303-2434},
	journal = {International Journal of Applied Earth Observation and Geoinformation},
	author = {Chen, Lin and Ren, Chunying and Zhang, Bai and Wang, Zongming and Liu, Mingyue and Man, Weidong and Liu, Jiafu},
	month = aug,
	year = {2021},
	pages = {102326}
}

@article{wang2019crop,
	title = {Crop type mapping without field-level labels: {Random} forest transfer and unsupervised clustering techniques},
	volume = {222},
	issn = {0034-4257},
	shorttitle = {Crop type mapping without field-level labels},
	journal = {Remote Sensing of Environment},
	author = {Wang, Sherrie and Azzari, George and Lobell, David B.},
	month = mar,
	year = {2019},
	pages = {303--317}
}

@article{kluger2021two,
	Author = {Dan M. Kluger and Sherrie Wang and David B. Lobell},
	Journal = {Remote Sensing of Environment},
	Pages = {112488},
	Title = {Two shifts for crop mapping: Leveraging aggregate crop statistics to improve satellite-based maps in new regions},
	Volume = {262},
	Year = {2021}}

@article{potapov2021mapping,
  title={Mapping global forest canopy height through integration of GEDI and Landsat data},
  author={Potapov, Peter and Li, Xinyuan and Hernandez-Serna, Andres and Tyukavina, Alexandra and Hansen, Matthew C and Kommareddy, Anil and Pickens, Amy and Turubanova, Svetlana and Tang, Hao and Silva, Carlos Edibaldo and others},
  journal={Remote Sensing of Environment},
  volume={253},
  pages={112165},
  year={2021},
  publisher={Elsevier}
}

@article{healey2020highly,
  title={Highly local model calibration with a new GEDI LiDAR asset on Google Earth Engine reduces landsat forest height signal saturation},
  author={Healey, Sean P and Yang, Zhiqiang and Gorelick, Noel and Ilyushchenko, Simon},
  journal={Remote Sensing},
  volume={12},
  number={17},
  pages={2840},
  year={2020},
  publisher={Multidisciplinary Digital Publishing Institute}
}

@article{qi2019forest,
  title={Forest biomass estimation over three distinct forest types using TanDEM-X InSAR data and simulated GEDI lidar data},
  author={Qi, Wenlu and Saarela, Svetlana and Armston, John and St{\aa}hl, G{\"o}ran and Dubayah, Ralph},
  journal={Remote Sensing of Environment},
  volume={232},
  pages={111283},
  year={2019},
  publisher={Elsevier}
}

@article{bruggisser2021potential,
  title={Potential of Sentinel-1 C-Band Time Series to Derive Structural Parameters of Temperate Deciduous Forests},
  author={Bruggisser, Moritz and Dorigo, Wouter and Dost{\'a}lov{\'a}, Alena and Hollaus, Markus and Navacchi, Claudio and Schlaffer, Stefan and Pfeifer, Norbert},
  journal={Remote Sensing},
  volume={13},
  number={4},
  pages={798},
  year={2021},
  publisher={Multidisciplinary Digital Publishing Institute}
}

@article{jin2019smallholder,
  title={Smallholder maize area and yield mapping at national scales with Google Earth Engine},
  author={Jin, Zhenong and Azzari, George and You, Calum and Di Tommaso, Stefania and Aston, Stephen and Burke, Marshall and Lobell, David B},
  journal={Remote sensing of environment},
  volume={228},
  pages={115--128},
  year={2019},
  publisher={Elsevier}
}

@article{nakalembe2021review,
  title={A review of satellite-based global agricultural monitoring systems available for Africa},
  author={Nakalembe, Catherine and Becker-Reshef, Inbal and Bonifacio, Rogerio and Hu, Guangxiao and Humber, Micheal Lawrence and Justice, Christina Jade and Keniston, John and Mwangi, Kenneth and Rembold, Felix and Shukla, Shraddhanand and others},
  journal={Global Food Security},
  volume={29},
  pages={100543},
  year={2021},
  publisher={Elsevier}
}

@article{soler2021maize,
  title={Maize and sorghum field segregation using multi-temporal Sentinel-2 data in central Mexico},
  author={Soler-P{\'e}rez-Salazar, Mar{\'\i}a J and Ortega-Garc{\'\i}a, Nicol{\'a}s and Vaca-Mier, Mabel and Cram-Hyedric, Silke},
  journal={Journal of Applied Remote Sensing},
  volume={15},
  number={2},
  pages={024513},
  year={2021},
  publisher={International Society for Optics and Photonics}
}

@article{veloso2017understanding,
  title={Understanding the temporal behavior of crops using Sentinel-1 and Sentinel-2-like data for agricultural applications},
  author={Veloso, Amanda and Mermoz, St{\'e}phane and Bouvet, Alexandre and Le Toan, Thuy and Planells, Milena and Dejoux, Jean-Fran{\c{c}}ois and Ceschia, Eric},
  journal={Remote Sensing of Environment},
  volume={199},
  pages={415--426},
  year={2017},
  publisher={Elsevier}
}

@article{belgiu2018sentinel,
  title={Sentinel-2 cropland mapping using pixel-based and object-based time-weighted dynamic time warping analysis},
  author={Belgiu, Mariana and Csillik, Ovidiu},
  journal={Remote Sensing of Environment},
  volume={204},
  pages={509--523},
  year={2018},
  publisher={Elsevier}
}

@misc{cdl,
title={{USDA National Agricultural Statistics Service Cropland Data Layer}. {Published crop-specific data layer [Online]}. {Available at https://nassgeodata.gmu.edu/CropScape/} (accessed {2021-07-01}; verified {2021-07-01})},
author={USDA-NASS},
year={2020},
institution={USDA-NASS, Washington, DC.}
}

@misc{cdl_metadata,
title={{2019 Iowa Cropland Data Layer [Online]}. {Available at https://www.nass.usda.gov/Research\_and\_Science/Cropland/metadata/metadata\_ia19.htm} (accessed {2021-07-01}; verified {2021-07-01})},
author={USDA-NASS},
year={2019},
institution={USDA-NASS, Washington, DC.}
}

@article{boryan2011monitoring,
author = {Claire   Boryan  and  Zhengwei   Yang  and  Rick   Mueller  and  Mike   Craig},
title = {Monitoring US agriculture: the US Department of Agriculture, National Agricultural Statistics Service, Cropland Data Layer Program},
journal = {Geocarto International},
volume = {26},
number = {5},
pages = {341-358},
year  = {2011},
publisher = {Taylor & Francis},
doi = {10.1080/10106049.2011.562309}
}

@misc{france,
title={{Registre parcellaire graphique (RPG): contours des parcelles et îlots culturaux et leur groupe de cultures majoritaire}. {Available at https://www.data.gouv.fr/en/datasets/registre-parcellaire-graphique-rpg-contours-des-parcelles-et-ilots-culturaux-et-leur-groupe-de-cultures-majoritaire/} (accessed {2021-07-01}; verified {2021-07-01})},
author={{Agence de Services et de Paiement}},
year={2019},
}

@techreport{asp2019rpg,
	Author = {{Agence de Services et de Paiement}},
	Institution = {{Institut National de l'Information Géographique et Forestière}},
	Title = {{RPG Version 2.0: Registre Parcellaire Graphique}},
	Year = {2019}}

@article{gitelson2005remote,
author = {Gitelson, Anatoly A. and Vina, Andres and Ciganda, Veronica and Rundquist, Donald C. and Arkebauer, Timothy J.},
title = {Remote estimation of canopy chlorophyll content in crops},
journal = {Geophysical Research Letters},
volume = {32},
number = {8},
pages = {},
year = {2005},
doi = {10.1029/2005GL022688}
}

@misc{agreste,
    title={Cultures developpees (hors fourrage, prairies, fruits, fleurs et vigne) [Online]. {Available at https://agreste.agriculture.gouv.fr/agreste-web/disaron/SAANR\_DEVELOPPE\_2/detail/} (accessed {2021-07-23}; verified {2021-07-23})},
    year={2019},
    author={{Le Service statistique minist\'eriel de l'agriculture}}
}

@misc{aafc,
title={{Annual Crop Inventory [Online]}. {Available at https://open.canada.ca/data/en/dataset/ba2645d5-4458-414d-b196-6303ac06c1c9} (accessed {2021-07-01}; verified {2021-07-01})},
author={{Agriculture and Agri-Food Canada}},
year={2021},
institution={Agriculture and Agri-Food Canada}
}

@article{defourny2019near,
	Author = {Pierre Defourny and Sophie Bontemps and Nicolas Bellemans and Cosmin Cara and G{\'e}rard Dedieu and Eric Guzzonato and Olivier Hagolle and Jordi Inglada and Laurentiu Nicola and Thierry Rabaute and Mickael Savinaud and Cosmin Udroiu and Silvia Valero and Agn{\`e}s B{\'e}gu{\'e} and Jean-Fran{\c c}ois Dejoux and Abderrazak {El Harti} and Jamal Ezzahar and Nataliia Kussul and Kamal Labbassi and Valentine Lebourgeois and Zhang Miao and Terrence Newby and Adolph Nyamugama and Norakhan Salh and Andrii Shelestov and Vincent Simonneaux and Pierre Sibiry Traore and Souleymane S. Traore and Benjamin Koetz},
	Journal = {Remote Sensing of Environment},
	Pages = {551-568},
	Title = {{Near real-time agriculture monitoring at national scale at parcel resolution: Performance assessment of the Sen2-Agri automated system in various cropping systems around the world}},
	Volume = {221},
	Year = {2019}}

@article{foerster2012crop,
	Author = {Saskia Foerster and Klaus Kaden and Michael Foerster and Sibylle Itzerott},
	Journal = {Computers and Electronics in Agriculture},
	Pages = {30-40},
	Title = {Crop type mapping using spectral--temporal profiles and phenological information},
	Volume = {89},
	Year = {2012}}

@article{jean2019tile2vec,
	Author = {Jean, Neal and Wang, Sherrie and Samar, Anshul and Azzari, George and Lobell, David and Ermon, Stefano},
	Journal = {Proceedings of the AAAI Conference on Artificial Intelligence},
	Month = {Jul.},
	Number = {01},
	Pages = {3967-3974},
	Title = {Tile2Vec: Unsupervised Representation Learning for Spatially Distributed Data},
	Volume = {33},
	Year = {2019}}

@InProceedings{rustowicz2019semantic,
    author = {Rustowicz, Rose and Cheong, Robin and Wang, Lijing and Ermon, Stefano and Burke, Marshall and Lobell, David},
    title = {Semantic Segmentation of Crop Type in Africa: A Novel Dataset and Analysis of Deep Learning Methods},
    booktitle = {Proceedings of the IEEE/CVF Conference on Computer Vision and Pattern Recognition (CVPR) Workshops},
    month = {June},
    year = {2019}
    }

@InProceedings{tseng2021learning,
    author    = {Tseng, Gabriel and Kerner, Hannah and Nakalembe, Catherine and Becker-Reshef, Inbal},
    title     = {Learning To Predict Crop Type From Heterogeneous Sparse Labels Using Meta-Learning},
    booktitle = {Proceedings of the IEEE/CVF Conference on Computer Vision and Pattern Recognition (CVPR) Workshops},
    month     = {June},
    year      = {2021},
    pages     = {1111-1120}
}

@article{wang2020mapping,
	Author = {Wang, Sherrie and Di Tommaso, Stefania and Faulkner, Joey and Friedel, Thomas and Kennepohl, Alexander and Strey, Rob and Lobell, David B.},
	Journal = {Remote Sensing},
	Number = {18},
	Title = {Mapping Crop Types in Southeast India with Smartphone Crowdsourcing and Deep Learning},
	Volume = {12},
	Year = {2020}}

@article{lambert2018estimating,
	Author = {Marie-Julie Lambert and Pierre C. Sibiry Traor{\'e} and Xavier Blaes and Philippe Baret and Pierre Defourny},
	Journal = {Remote Sensing of Environment},
	Pages = {647-657},
	Title = {Estimating smallholder crops production at village level from Sentinel-2 time series in Mali's cotton belt},
	Volume = {216},
	Year = {2018}}

@article{belgiu2021phenology,
	Author = {Mariana Belgiu and Wietske Bijker and Ovidiu Csillik and Alfred Stein},
	Journal = {International Journal of Applied Earth Observation and Geoinformation},
	Pages = {102264},
	Title = {Phenology-based sample generation for supervised crop type classification},
	Volume = {95},
	Year = {2021}}

@article{hofton2000decomposition,
  title={Decomposition of laser altimeter waveforms},
  author={Hofton, Michelle A and Minster, Jean-Bernard and Blair, J Bryan},
  journal={IEEE Transactions on geoscience and remote sensing},
  volume={38},
  number={4},
  pages={1989--1996},
  year={2000},
  publisher={IEEE}
}

@INPROCEEDINGS{ho1995random,
  author={Tin Kam Ho},
  booktitle={Proceedings of 3rd International Conference on Document Analysis and Recognition}, 
  title={Random decision forests}, 
  year={1995},
  volume={1},
  number={},
  pages={278-282 vol.1},
  doi={10.1109/ICDAR.1995.598994}}

@article{breiman2001random,
	Author = {Breiman, Leo},
	Journal = {Machine Learning},
	Number = {1},
	Pages = {5--32},
	Title = {Random Forests},
	Volume = {45},
	Year = {2001}}

@article{waldner2020deep,
  title={Deep learning on edge: Extracting field boundaries from satellite images with a convolutional neural network},
  author={Waldner, Fran{\c{c}}ois and Diakogiannis, Foivos I},
  journal={Remote Sensing of Environment},
  volume={245},
  pages={111741},
  year={2020},
  publisher={Elsevier}
}
